\documentclass{ckm}
\usepackage{txfonts}

\confname{Workshop on the CKM Unitarity Triangle, IPPP Durham, April 2003}

\title{Prospects for $B_{d,s}\rightarrow h^+h^-$ measurements at LHCb}

\author{V. M. Vagnoni}
\address{Dipartimento di Fisica di Bologna and INFN, Sezione di Bologna,
via Irnerio 46, I-40126 Bologna, Italy}

\begin{document}

\begin{abstract}
LHCb will collect large samples of $B_{d}$ and $B_{s}$ decays. Combining the
CP-violating observables of the decays $B_{d}\rightarrow\pi^{+}\pi^{-}$
and $B_{s}\rightarrow K^{+}K^{-}$ it is possible to extract the
$\gamma$ angle of the unitarity triangle. The selection of these
decays within the current LHCb simulation framework is outlined and the
expected annual event yields and background-to-signal ratios are
quoted. Then, the results of a study on the sensitivity that LHCb can achieve
for the corresponding CP-violating observables are presented.
\end{abstract}
\maketitle

\section{Introduction}

\begin{figure}
\begin{center}\includegraphics[
width=1.1\columnwidth]{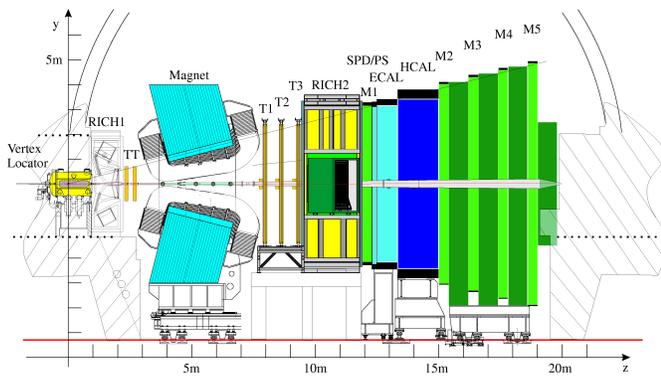}\end{center}

\caption{\label{detector}Schematic drawing of the LHCb detector.}
\end{figure}

LHCb is a dedicated experiment on b-quark physics, currently under
construction at the Large Hadron Collider (LHC). It will profit from
the 500 $\mu b$ b-hadron production cross section in the 14 TeV proton-proton
collisions at LHC (i.e. about 1\% of the total visible cross section)
to make precise measurements of CP-violation and rare decays of the
B-mesons \cite{LHCbTP}. By over-constraining the Cabibbo-Kobayashi-Maskawa
matrix elements, LHCb will hopefully be able to observe subtle inconsistencies
with the Standard Model, therefore providing indications of new physics.
A sketch of the LHCb detector is shown in Fig. \ref{detector}.

Profiting from the large $b\overline{b}$ cross section, LHCb will
collect large samples of $B_{d}$ and $B_{s}$ decays. Combining the
CP-violating observables of the decays $B_{d}\rightarrow\pi^{+}\pi^{-}$
and $B_{s}\rightarrow K^{+}K^{-}$ it is possible, by means of a method
proposed by R. Fleischer \cite{Fleischer}, to extract the $\gamma$ angle of
the unitarity triangle.

In section 2 the selection strategy, implemented to identify the
$B_{d}\rightarrow\pi^{+}\pi^{-}$
and $B_{s}\rightarrow K^{+}K^{-}$ decays within the current LHCb
Monte Carlo simulation framework, is described and the estimated annual event
yields and background levels are quoted.

By using the results shown in section 2, the sensitivities achievable
on the CP-violating observables can be determined. This is described
in section 3.

\section{Event reconstruction}

To evaluate the performance of LHCb in the reconstruction
of $B_{d}\rightarrow\pi^{+}\pi^{-}$ and $B_{s}\rightarrow K^{+}K^{-}$
decays, a full GEANT Monte Carlo simulation has been performed, followed
by realistic pattern recognition algorithms for the track reconstruction
and particle identification. The Monte Carlo event samples used for
this analysis are shown in Tab. \ref{mc samples}.

\begin{table}
\begin{center}\begin{tabular}{|c|c|}
\hline 
Channel&
Monte Carlo Statistics\tabularnewline
\hline
\hline 
$B_{d}\rightarrow\pi^{+}\pi^{-}$&
60000\tabularnewline
\hline 
$B_{s}\rightarrow K^{+}K^{-}$&
65000\tabularnewline
\hline 
$B_{d}\rightarrow K^{+}\pi^{-}$&
62000\tabularnewline
\hline 
$B_{s}\rightarrow K^{-}\pi^{+}$&
24000\tabularnewline
\hline 
$\Lambda_{b}\rightarrow pK^{-}$&
20000\tabularnewline
\hline 
$\Lambda_{b}\rightarrow p\pi^{-}$&
21000\tabularnewline
\hline
$b\bar{b}\rightarrow X$&
1144000\tabularnewline
\hline
\end{tabular}\end{center}

\caption{\label{mc samples}Monte Carlo event samples used to study the selection
of $B_{d}\rightarrow\pi^{+}\pi^{-}$ and $B_{s}\rightarrow K^{+}K^{-}$
decays. The other channels are studied as potential sources of background.}
\end{table}

Two sources of background have been considered: \emph{two charged
body decays} of B mesons and baryons, which can fake the signal in
case of particle mis-identification, and combinatorial background.
The dominant source of combinatorial background is believed to come
from \emph{beauty} events.

To select the interesting decays and reject the backgrounds, a set
of kinematical and topological cuts has been used. For each pair
of tracks with opposite
charge identified as pions or kaons by the RICH detectors, cuts are
applied on:

\begin{itemize}
\item momentum;
\item smallest and largest transverse momentum;
\item smallest and largest impact parameter significance;
\item $\chi^{2}$ of common vertex fit.
\end{itemize}
Each pair surviving to these cuts is used to form a B-meson candidate,
and further cuts are applied on it:

\begin{itemize}
\item transverse momentum;
\item impact parameter significance;
\item distance of flight significance;
\item invariant mass.
\end{itemize}
With these set of cuts, with values specifically optimized for each of the two
decays under study, all the combinatorial $b\bar{b}$ background events,
in the limited Monte Carlo statistics available at the moment of this
analysis, are rejected. At the same time the two body B decays studied
as specific backgrounds, thanks to the particle identification performance
relying on the two RICH detectors \cite{RICH}, are maintained under control.

\begin{figure}
\begin{center}\includegraphics[%
  width=0.8\columnwidth]{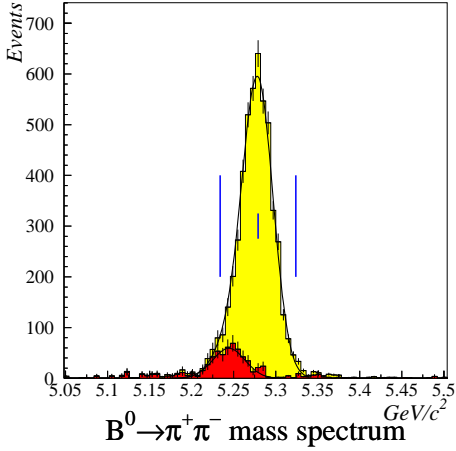}\end{center}

\caption{\label{mass}Invariant mass distribution for $B_{d}\rightarrow\pi^{+}\pi^{-}$
reconstructed events, together with the mass spectrum due to other
two body B-decays background. The mass resolution is about $18\, MeV/c^{2}$. }
\end{figure}

Fig. \ref{mass} shows the invariant mass distribution for the reconstructed
$B_{d}\rightarrow\pi^{+}\pi^{-}$ events surviving the selection, together
with the two body B-decay background. The estimated annual event yields
and background-to-signal ratios can be read in Tab. \ref{yields}.

\begin{table}
\begin{center}\begin{tabular}{|c|c|c|}
\hline 
Channel&
B/S&
Untagged annual yield\tabularnewline
\hline
\hline 
$B_{d}\rightarrow\pi^{+}\pi^{-}$&
$<0.8$&
$27000$\tabularnewline
\hline 
$B_{s}\rightarrow K^{+}K^{-}$&
$<0.55$&
$35000$\tabularnewline
\hline
\end{tabular}\end{center}

\caption{\label{yields}Estimated number of offline-selected events per year
and background-to-signal ratios for the $B_{d}\rightarrow\pi^{+}\pi^{-}$
and $B_{s}\rightarrow K^{+}K^{-}$ decays at LHCb. }
\end{table}

\section{CP sensitivity}

The time dependent CP asymmetry takes the usual form:

\[
\mathcal{A}_{CP}(\tau)=\mathcal{A}_{CP}^{dir}\cdot cos(x\cdot\tau)+\mathcal{A}_{CP}^{mix}sin(x\cdot\tau)\]
which is valid assuming the decay width difference of the B mass eigenstates
$\Delta\Gamma$ to be negligible.

In order to generate the requested samples of tagged $B_{d}\rightarrow\pi^{+}\pi^{-}$
and $B_{s}\rightarrow K^{+}K^{-}$ decays to perform a study on the
sensitivity that LHCb can achieve on the measurements of $\mathcal{A}_{CP}^{dir}$
and $\mathcal{A}_{CP}^{mix}$, a standalone toy Monte Carlo program has been
used. The program takes as input the number of reconstructed events
per year and the background levels shown in Tab. \ref{yields}. It reproduces
the acceptance as a function of the proper time and the proper time
resolution as studied in detail by using the full GEANT simulation and simulates
the effect of CP violation using given values of $\mathcal{A}_{CP}^{dir}$ and
$\mathcal{A}_{CP}^{mix}$, then simulates the tagging procedure,
finally giving as output a sample of proper times of tagged $B$ and $\bar{B}$ decays.

To {}``measure'' from this sample of proper times the values of
$\mathcal{A}_{CP}^{dir}$ and $\mathcal{A}_{CP}^{mix}$, with their uncertainties and
correlation, an unbinned maximum likelihood method has
been used. The log-likelihood function to be maximised with respect
to $\mathcal{A}_{CP}^{dir}$ and $\mathcal{A}_{CP}^{mix}$ is:

\[
\log \mathcal{L}=\log\left[\prod_{i}n(\tau_{i})\cdot\prod_{j}\bar{n}(\tau_{j})\right]\]
where \{$\tau_{i}$\} and \{$\tau_{j}$\} are the sets of decay proper
times for the tagged $B$ and $\bar{B}$ respectively, and

\[
n(\tau_{i})=\left[e^{-t/\tau_{B}}\cdot\left(1+\frac{1-2\omega}{1+B/S}\cdot \mathcal{A}_{CP}(t)\right)\cdot\epsilon(t)\right]\bigotimes R(\tau_{i}-t)\]

\[
\bar{n}(\tau_{j})=\left[e^{-t/\tau_{B}}\cdot\left(1-\frac{1-2\omega}{1+B/S}\cdot \mathcal{A}_{CP}(t)\right)\cdot\epsilon(t)\right]\bigotimes R(\tau_{j}-t)\]
being $\tau_{B}$ the B-meson lifetime, B/S the background-to-signal ratio,
$\omega$ the wrong tagging
fraction, $\epsilon(t)$ the acceptance as function of the
proper time, $R(\tau-t)$ a function accounting for the proper time
resolution of the LHCb spectrometer (normal distribution with a width
of $40\, fs$) and where the symbol $\bigotimes$ stands for convolution
product.

\begin{figure}
\begin{center}\begin{tabular}{c}
\includegraphics[%
  width=0.6\columnwidth]{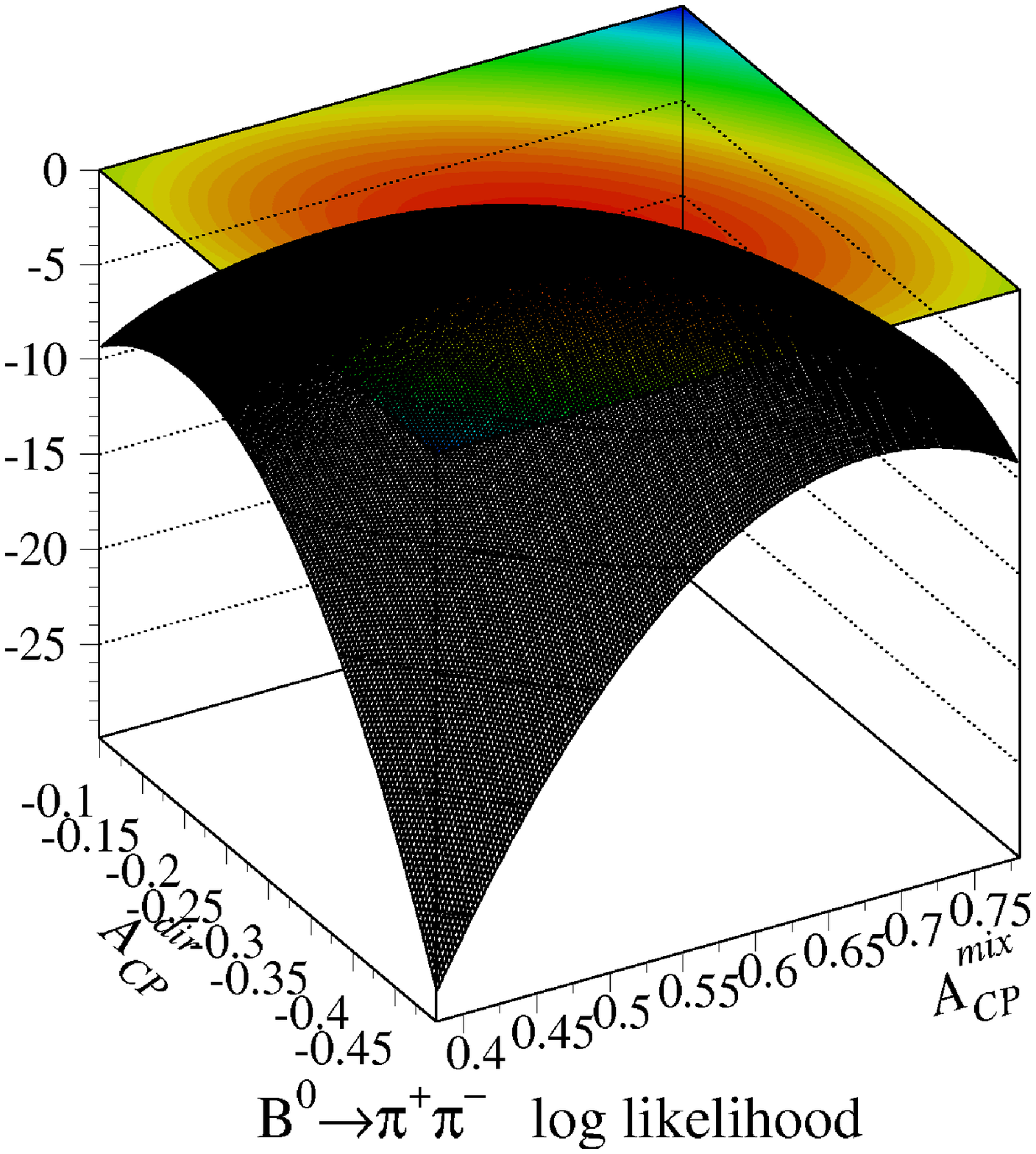}\tabularnewline
\includegraphics[%
  width=0.6\columnwidth]{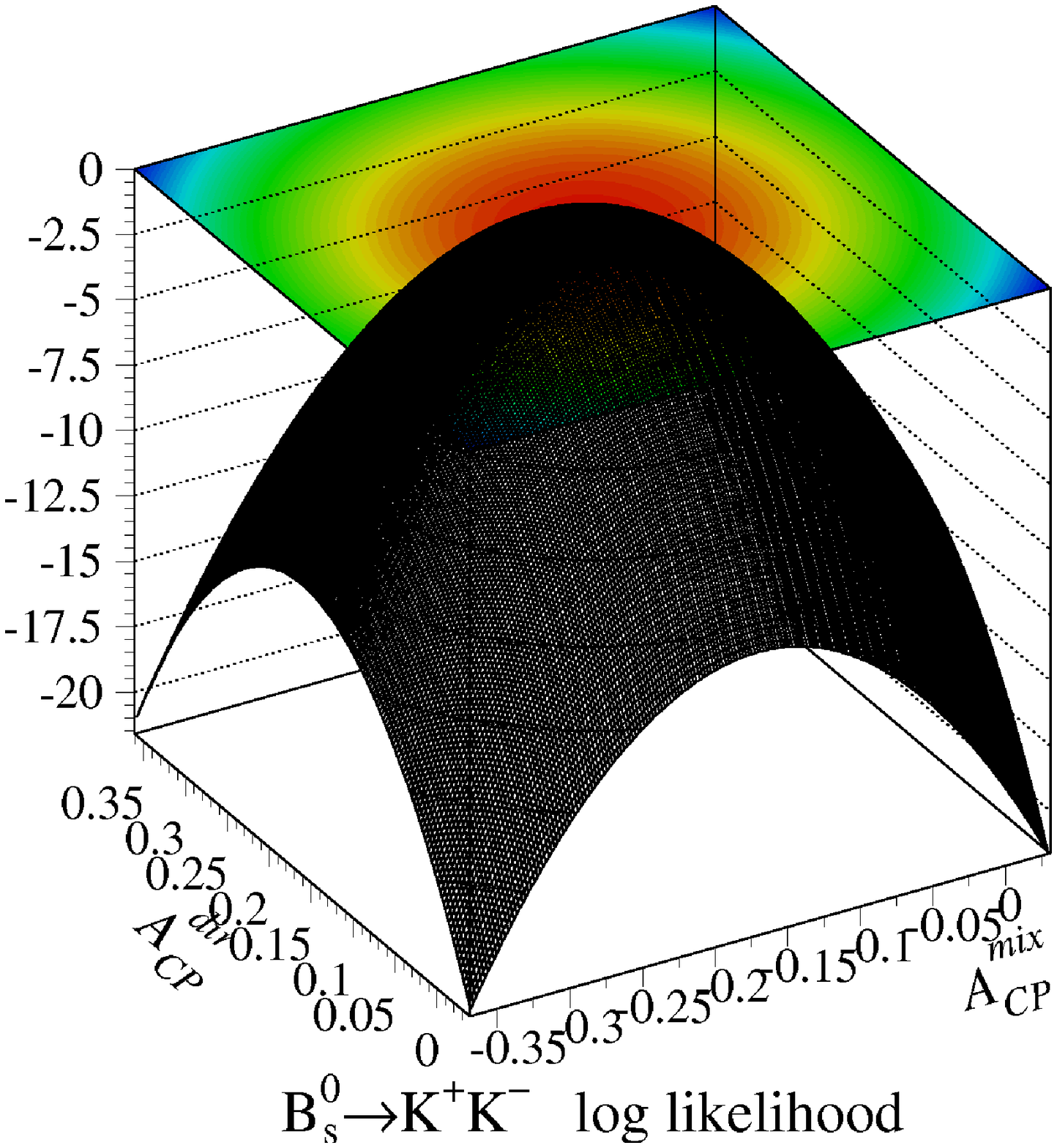}\tabularnewline
\end{tabular}\end{center}

\caption{\label{loglikelihood}Dependence of the log-likelihood on $\mathcal{A}_{CP}^{dir}$
and $\mathcal{A}_{CP}^{mix}$ for the $B_{d}\rightarrow\pi^{+}\pi^{-}$ decay
(upper plot) and for the $B_{s}\rightarrow K^{+}K^{-}$ decay (lower
plot).}
\end{figure}

\begin{figure}
\begin{center}\begin{tabular}{c}
\includegraphics[%
  width=0.6\columnwidth]{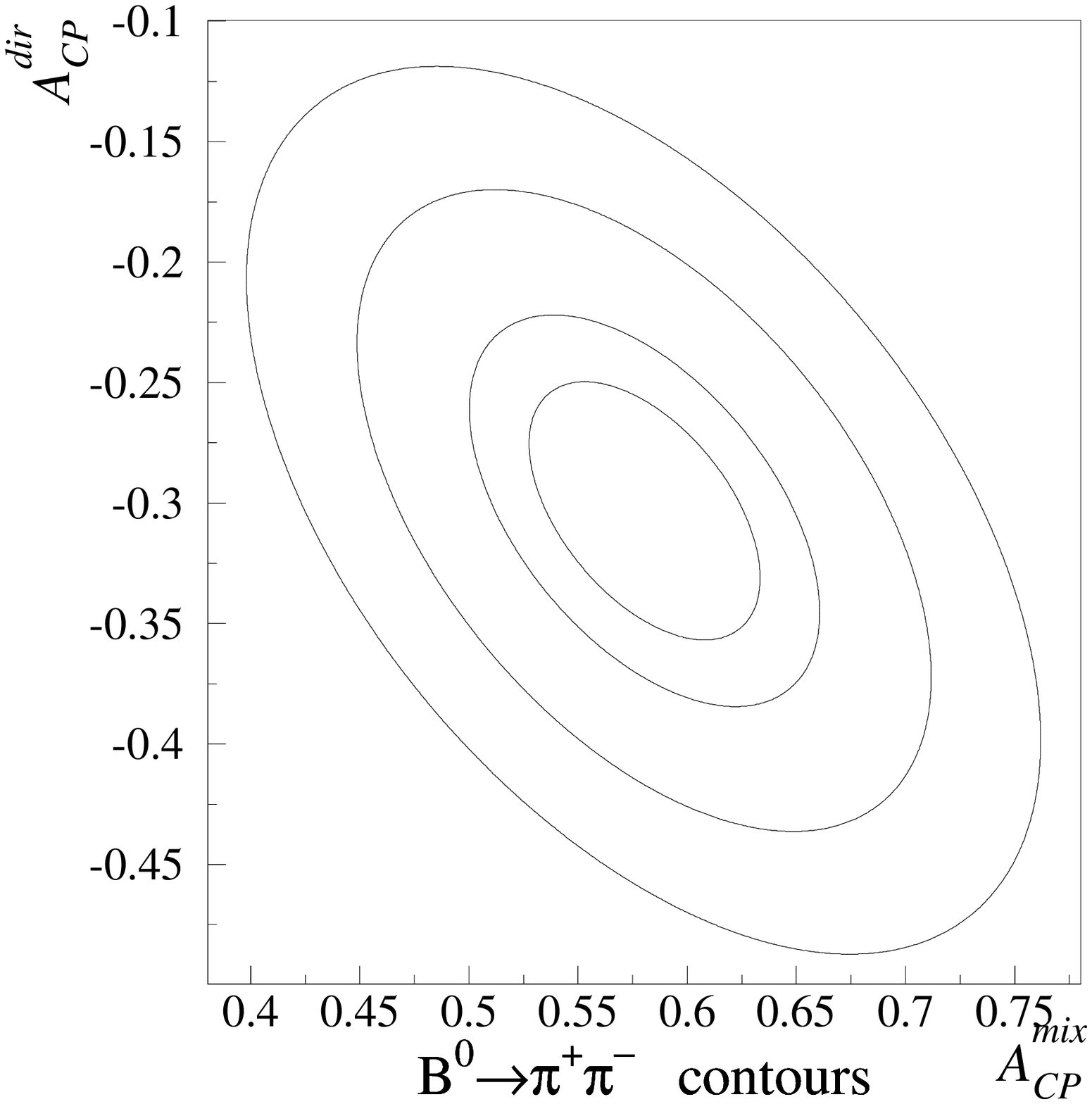}\tabularnewline
\includegraphics[%
  width=0.6\columnwidth]{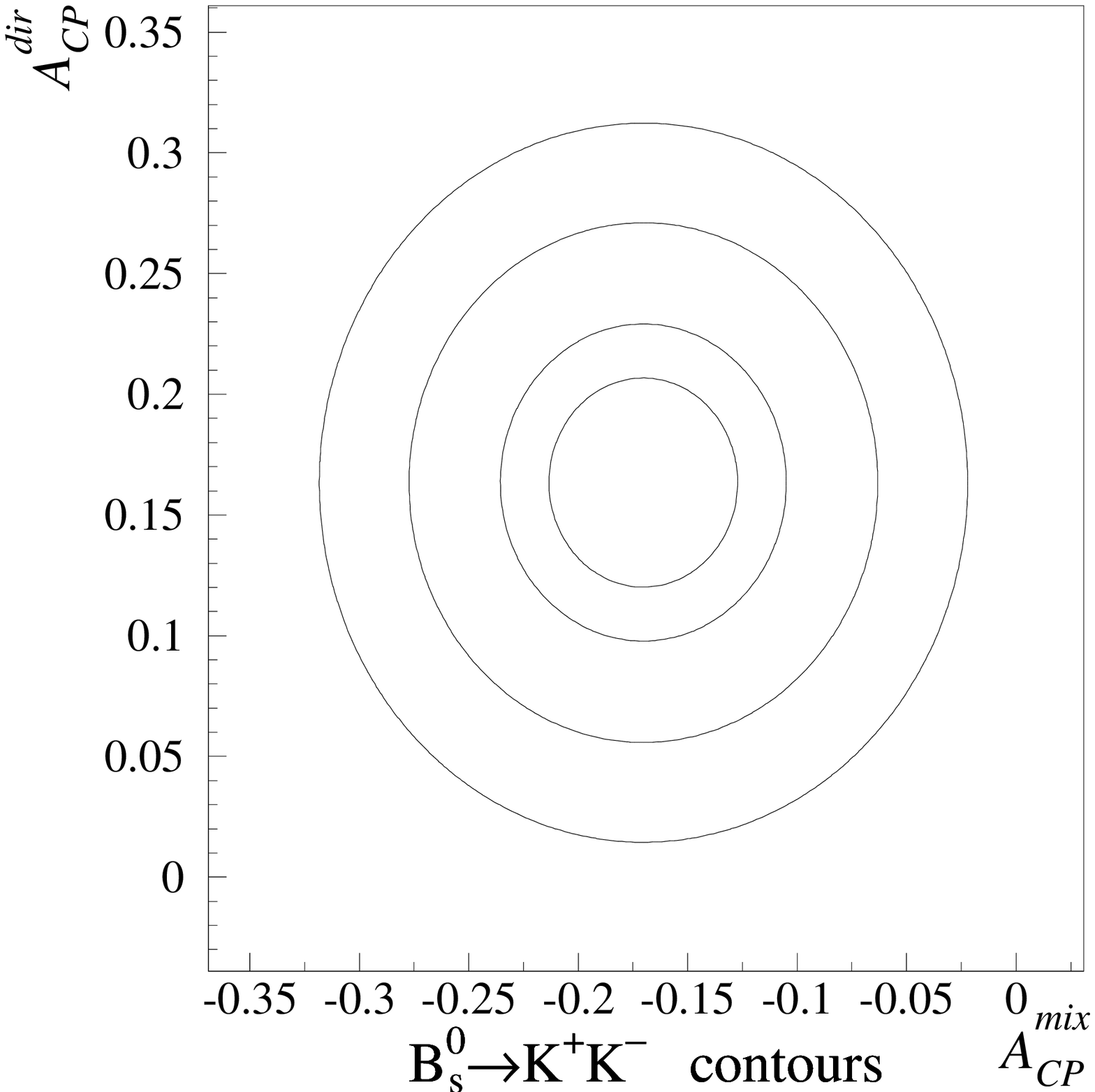}\tabularnewline
\end{tabular}\end{center}

\caption{\label{confidence}Confidence regions for $\mathcal{A}_{CP}^{dir}$ and $\mathcal{A}_{CP}^{mix}$
for the $B_{d}\rightarrow\pi^{+}\pi^{-}$ decay (upper plot) and for
the $B_{s}\rightarrow K^{+}K^{-}$ decay (lower plot), corresponding
to one year of LHCb data taking. The four ellipses correspond to the
39\% (standard error ellipse), 68\%, 95\% and 99\% confidence regions.}
\end{figure}

\begin{figure}
\begin{center}
\includegraphics[%
  width=0.8\columnwidth]{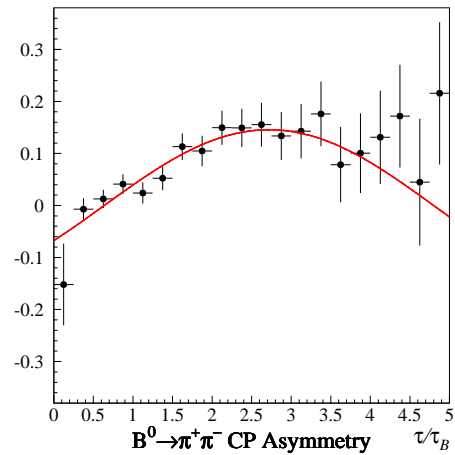}
\end{center}

\caption{\label{asymmetry}Binned asymmetry for the $B_{d}\rightarrow\pi^{+}\pi^{-}$
decay with the result of the likelihood fit superimposed.}
\end{figure}

As an example, Fig. \ref{loglikelihood} shows the dependence of $\log \mathcal{L}$
on $\mathcal{A}_{CP}^{dir}$ and $\mathcal{A}_{CP}^{mix}$ for a sample of $B_{d}\rightarrow\pi^{+}\pi^{-}$
and a sample of $B_{s}\rightarrow K^{+}K^{-}$ events generated by the
toy Monte Carlo program, corresponding to one year of LHCb data taking,
where the following input values for the CP asymmetries have been
used: $\mathcal{A}_{CP}^{dir}(B_{d}\rightarrow\pi^{+}\pi^{-})=-0.3$, $\mathcal{A}_{CP}^{mix}(B_{d}\rightarrow\pi^{+}\pi^{-})=0.58$,
$\mathcal{A}_{CP}^{dir}(B_{s}\rightarrow K^{+}K^{-})=0.16$ and $\mathcal{A}_{CP}^{mix}(B_{s}\rightarrow K^{+}K^{-})=-0.17$.
Fig. \ref{confidence} shows the corresponding confidence regions
determined for $\mathcal{A}_{CP}^{dir}$ and $\mathcal{A}_{CP}^{mix}$, while Fig. \ref{asymmetry}
shows the binned asymmetry for the $B_{d}\rightarrow\pi^{+}\pi^{-}$
decay with the result of the likelihood fit superimposed.

The obtained resolutions and correlations for $\mathcal{A}_{CP}^{dir}$ and
$\mathcal{A}_{CP}^{mix}$, corresponding to one year of LHCb data taking, are
shown in Tab. \ref{results}. The measurements of $\mathcal{A}_{CP}^{dir}$
and $\mathcal{A}_{CP}^{mix}$ are strongly correlated 
in the case of the $B_{d}\rightarrow\pi^{+}\pi^{-}$
channel, while the correlation for the $B_{s}\rightarrow K^{+}K^{-}$
channel is negligible, as it can easily be argued by looking at the
contour plots of Fig. \ref{confidence}.

\begin{table}
\begin{center}\begin{tabular}{|c|c|c|c|}
\hline 
Channel&
$\sigma(\mathcal{A}_{CP}^{dir})$&
$\sigma(\mathcal{A}_{CP}^{mix})$&
$Corr(\mathcal{A}_{CP}^{dir},\, \mathcal{A}_{CP}^{mix})$\tabularnewline
\hline
\hline 
$B_{d}\rightarrow\pi^{+}\pi^{-}$&
0.054&
0.054&
-0.53\tabularnewline
\hline 
$B_{s}\rightarrow K^{+}K^{-}$&
0.043&
0.043&
0\tabularnewline
\hline
\end{tabular}\end{center}

\caption{\label{results}Resolutions on the measurements of $\mathcal{A}_{CP}^{dir}$
and $\mathcal{A}_{CP}^{mix}$ for the $B_{d}\rightarrow\pi^{+}\pi^{-}$ and
$B_{s}\rightarrow K^{+}K^{-}$ decays expected at LHCb after one year of
data taking. The last column reports the correlation between the measured
values of the asymmetries.}
\end{table}

\begin{figure}
\begin{center}\includegraphics[%
  width=0.8\columnwidth]{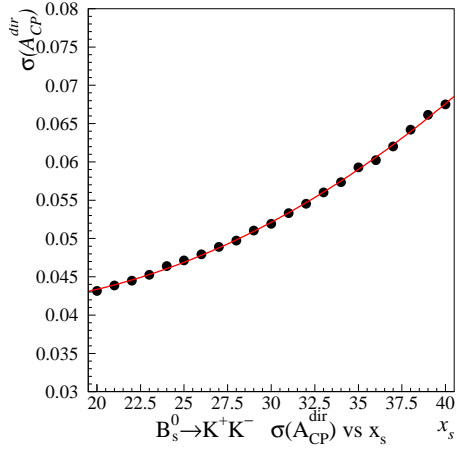}\end{center}

\caption{\label{xsstudy}Dependence of the error on $\mathcal{A}_{CP}^{dir}$ for the
$B_{s}\rightarrow K^{+}K^{-}$ decay on the value of the mixing parameter $x_s$.}
\end{figure}

The previous results for the $B_{s}\rightarrow K^{+}K^{-}$ channel
are obtained for a value of the mixing parameter $x_{s}=20$.
Due to the finite proper time
resolution of the spectrometer (about $40\, fs$), higher values of
$x_{s}$, i.e. faster $B_{s}-\bar{B}_{s}$ oscillations, lead to larger
errors for the measurements
of $\mathcal{A}_{CP}^{dir}$ and $\mathcal{A}_{CP}^{mix}$.
Thus, a study of the dependence of the resolution of $\mathcal{A}_{CP}^{dir}$
and $\mathcal{A}_{CP}^{mix}$ on $x_{s}$, up to $x_{s}=40$, has been performed.
The results of this study for $\mathcal{A}_{CP}^{dir}$ are shown in Fig. \ref{xsstudy}
(the dependence for $\mathcal{A}_{CP}^{mix}$ is identical and is omitted here).
As it can be seen the error on the asymmetries increases by about
a factor 1.6 at $x_{s}=40$ with respect to $x_{s}=20$.

\section{Conclusions}
The results presented in this paper show the potential of LHCb
in collecting large samples of $B_{d}\rightarrow \pi^{+}\pi^{-}$
and $B_{s}\rightarrow K^{+}K^{-}$ decays. They are obtained by using a full and realistic
Monte Carlo simulation of the detector response.

The number of $B_{d}\rightarrow \pi^{+}\pi^{-}$ offline-selected decays are
27000/year, while for $B_{s}\rightarrow K^{+}K^{-}$ it is 35000/year, with
background-to-signal ratios respectively of about 0.8 and 0.55.

The sensitivity on the CP-violating observables achievable by LHCb for these
channels has been studied in detail, yielding very encouraging results. The
estimated precision on $\mathcal{A}_{CP}^{dir}$ and $\mathcal{A}_{CP}^{mix}$ after one year
of data taking, assuming $x_{s}$ to be equal to 20,
is 0.054 for the $B_{d}\rightarrow \pi^{+}\pi^{-}$ channel and 0.043 for the
$B_{s}\rightarrow K^{+}K^{-}$ channel.

By means of these observables it is possible to extract the $\gamma$ angle of
the unitarity triangle. In fact, $\mathcal{A}_{CP}^{dir}$
and $\mathcal{A}_{CP}^{mix}$, related to the $B_{d}\rightarrow\pi^{+}\pi^{-}$
decay, can be expressed in the framework of the Standard Model as:

\[
\mathcal{A}_{CP}^{dir}(B_{d})=-\frac{2d\sin\theta\sin\gamma}{1-2d\cos\theta\cos\gamma+d^{2}}\]

\[
\mathcal{A}_{CP}^{mix}(B_{d})=\frac{\sin(\phi_{d}+2\gamma)-2d\cos\theta\sin(\phi_{d}+\gamma)+d^{2}\sin\phi_{d}}{1-2d\cos\theta\cos\gamma+d^{2}}\]
where $\phi_{d}=2\beta$ is the $B_{d}-\bar{B}_{d}$ mixing phase,
while $d$ and $\theta$
parametrize - sloppily speaking - the penguin over tree amplitude
ratio of the decay transition \cite{Fleischer}. Analogously,
$\mathcal{A}_{CP}^{dir}$ and $\mathcal{A}_{CP}^{mix}$, related to
the $B_{s}\rightarrow K^{+}K^{-}$ decay, can be written as

\[
\mathcal{A}_{CP}^{dir}(B_{s})=\frac{2\widetilde{d}{}^{\prime}\sin\theta^{\prime}\sin\gamma}{1+2\widetilde{d}{}^{\prime}\cos\theta^{\prime}\cos\gamma+\widetilde{d}{}^{\prime2}}\]

\[
\mathcal{A}_{CP}^{mix}(B_{s})=\frac{\sin(\phi_{s}+2\gamma)+2\widetilde{d}{}^{\prime}\cos\theta{}^{\prime}\sin(\phi_{s}+\gamma)+\widetilde{d}{}^{\prime2}\sin\phi_{s}}{1+2\widetilde{d}{}^{\prime}\cos\theta{}^{\prime}\cos\gamma+\widetilde{d}{}^{\prime2}}\]
where $\phi_{s}=2\delta\gamma$ is the $B_{s}-\bar{B}_{s}$ mixing
phase, $\widetilde{d}{}^{\prime}=\frac{1-\left|V_{us}\right|^{2}}{\left|V_{us}\right|}\, d^{\prime}$,
and the $d^{\prime}$ and $\theta{}^{\prime}$ parameters are analogous
to $d$ and $\theta$ for the $B_{s}\rightarrow K^{+}K^{-}$ transition.

In the limit of exact U-spin symmetry of the strong interactions the
relations $d=d'$ and $\theta=\theta'$ hold, and the measurements
of the four asymmetry coefficients allow to determine $\phi_{d}$
and $\gamma$ simultaneously, provided that $\phi_{s}$ is determined
elsewhere (e.g. through the $B_{s}\rightarrow J/\psi\phi$ transition)
or considered negligibly small, as it is expected in the Standard
Model. Moreover $\phi_{d}$ will be accurately known at the time of
LHCb, thus allowing a more precise determination of $\gamma$.

The resolution on $\gamma$ achievable at LHCb by using this method,
in case of negligible U-spin breaking effects, is of the order of few degrees.

\section*{Acknowledgements}
The author is grateful to T. Nakada and O. Schneider for their encouragement and
suggestions. Special thanks to U. Marconi for his help and practical support.

\end{document}